\documentclass[a4paper,12pt,english]{article}

\usepackage{amsfonts,bm,amssymb,euscript,array,babel,cite}
\usepackage{mathtools}
\usepackage{tikz-cd}
\usepackage{amsmath}
\usepackage{tensor}
\usepackage{hhline}
\usepackage[amsmath]{empheq}
\usepackage{hyperref,multirow}
\usepackage{cancel}
\usepackage{mathrsfs}
\usepackage{pdfsync}
\usepackage{tikz}

\usepackage{framed}

\newsavebox{\mysaveboxM}
\newsavebox{\mysaveboxT}

\makeatletter

\newcommand\gh{\operatorname{gh}}
\newcommand{\dd}{\mathrm{d}}
\newcommand{\DD}{\mathrm{D}}

\newcommand{\w}{\wedge}

\newcommand{\be}{\begin{equation}}
\newcommand{\ee}{\end{equation}}

\def\nn{\nonumber}
\def \bea{\begin{eqnarray}} 
\def\eea{\end{eqnarray}}
\newcommand{\mf}{\mathfrak}
\def\mc{\mathcal}

\def\bi{\begin{itemize}} 
\def\ei{\end{itemize}}

\makeatother

\def\a{\alpha}  \def\g{\gamma}  \def\d{\delta}

  \def\m{\mu}
    \def\r{\rho}
 \def\S{\Sigma}

\def\one{\mbox{1 \kern-.59em {\rm l}}}

\numberwithin{equation}{section}

\begin{document}

\makeatother
\parindent=0cm
\renewcommand{\title}[1]{\vspace{10mm}\noindent{\Large{\bf #1}}\vspace{8mm}} \newcommand{\authors}[1]{\noindent{\large #1}\vspace{5mm}} \newcommand{\address}[1]{{\itshape #1\vspace{2mm}}}

\begin{titlepage}

\begin{flushright}
	RBI-ThPhys-2022-18
\end{flushright}

\begin{center}

\title{ {\large {Dirac sigma models from gauging the nonlinear sigma models and its BV action}}}

  \authors{Grgur \v{S}imuni\'c{$^{\a,}$}{\footnote{Grgur.Simunic@irb.hr}}
    }
 
 \vskip 3mm
 
  \address{ $^{\a}$ Division of Theoretical Physics, Rudjer Bo\v skovi\'c Institute \\ Bijeni\v cka 54, 10000 Zagreb, Croatia \\

 }

\vskip 2cm

\begin{abstract}

We present the construction of the Dirac sigma models by gauging the 2-dimensional nonlinear sigma models, but also including the possibility of nonminimal coupling to the metric sector. We show that for a large variety of possible cases, the minimal coupling to the metric sector is the only nontrivial possibility. In addition, we present the construction of the classical Batalin-Vilkovisky action for the Dirac sigma models. We follow a direct approach in its construction, by requiring it to be a solution of the classical master equation.

\end{abstract}

\end{center}

\vskip 2cm

\end{titlepage}

\setcounter{footnote}{0}
\tableofcontents


\section{Introduction}
\label{sec1}

Two-dimensional sigma models play an important role in various physical situations. There is a whole subset of those theories, referred to as topological field theories, whose moduli space of classical solutions is finite-dimensional, up to gauge transformations. Some better known examples of such theories include A/B models \cite{Witten:1988xj, Witten:1991zz}, the G/G Wess-Zumino-Witten (WZW) model \cite{Alekseev:1995py} and the Poisson sigma model \cite{SchallerStrobl, Ikeda}. Furthermore, the research about the relation between WZW models and the Poisson sigma model \cite{Kotov:2004wz} led to the construction of the Dirac sigma models. These are specific 2-dimensional sigma models whose underlying structure is that of Dirac manifolds \cite{dirac}. They are also related to Dirac structures which are Lie algebroids obtained as maximal isotropic and involutive subbundles of an exact Courant algebroid \cite{Salnikov:2013pwa, ChatzistavrakidisAHP}. Special case of the Dirac sigma models is the Poisson sigma model, which is obtained when one considers a cotangent bundle as the Dirac structure. Since in general the Dirac sigma model include the Wess-Zumino term in the action described by a 3-form $H$, and since the Poisson sigma model is obtained when H vanishes, the Poisson sigma model can be generalized to $H$-twisted Poisson sigma model \cite{Klimcik:2001vg}, whose underlying structure is that of a twisted Poisson manifold \cite{SeveraWeinstein} instead of Poisson manifold.

The Poisson sigma model can be obtained in another way, through the AKSZ construction for the construction of the BV action \cite{Alexandrov:1995kv}. This is a geometric approach to the Batalin-Vilkovisky (BV) quantization of gauge theories. The AKSZ construction relies on the underlying structure of the QP$n$ manifold and the Poisson sigma model emerges as the $n=1$ case. However, for the twisted Poisson sigma model, and the Dirac sigma models in general, the QP$n$ does not necessarily exist. While one can always construct a Q-structure (a homological vector field), a (graded symplectic) P-structure does not have to exist. This obstruction is due to the 3-form $H$. This means that, as long as $H$ does not vanish, the AKSZ construction cannot be used, meaning other methods have to be used, like it has been done in \cite{Ikeda:2019czt} for the twisted Poisson sigma model.

To determine the BV action for the Dirac sigma models, we take a more traditional approach. First introducing ghosts (and since the theories here are irreducible, there is no need for ghosts for ghost), we define the BRST operator for all the fields. It turns out that the BRST operator is nilpotent only on-shell which means that the antifields have to be introduce in order to quantize these theories. Then the BV action $S_{BV}$ is constructed. Part of it is known directly from the BRST operator, while the rest is determined such that is satisfies the classical master equation:
\begin{equation}
(S_{BV},S_{BV})=0\,,
\end{equation}
where $(\cdot,\cdot)$ is the antibracket in the space of fields. We show that the extra contributions, ones that do not come from the BRST operator, are all quadratic in the antifields. Furthermore, the factors in those terms turn out to be basic curvatures of the two connections that control gauge transformations of the Dirac sigma model.

\section{Dirac Sigma models}
\label{sec2}

\subsection{Generalised gauging}
\label{sec21}

Propagation of strings in some $n$-dimensional spacetime $M$ is described by nonlinear sigma models, which are two-dimensional field theories on a worldsheet $\Sigma_2$. The main fields here are scalar fields $X^1,\ldots,X^n$ which are components of an embbeding function $X:\Sigma_2\rightarrow M$. This fields couple nonlinearly to the background fields of the target space. In the simplest case of the bosonic strings, those include the metric $g_{ij}$ and the Kalb-Ramond field $B_{ij}$, or more generally its curvature $H=\dd B$.\footnote{Also, there is a scalar dilaton, which will not be considered in this paper.} Here $H$ does not have to equal $\dd B$ globally, but only locally, meaning that $H$ is a closed form, but not necessarily exact. In terms of these 3-form, the sigma model action contains the Wess-Zumino term that exist not on $\Sigma_2$, but on a 3-dimensional manifold $\Sigma_3$ whose boundary equals $\Sigma_2$. For this reason, the embbeding function $X$ has to be extended in its domain to include $\Sigma_3$. Now the action functional takes the form:
\begin{equation}
S[X]=-\int_{\Sigma_2} \frac{1}{2}g_{ij}(X)\dd X^i\wedge \ast\dd X^j-\int_{\Sigma_3}\frac{1}{3!}H_{ijk}(X)\dd X^i\wedge\dd X^j\wedge\dd X^k\,,
\label{NLSM}
\end{equation}
where $g_{ij}(X)=X^\ast g_{ij}(x)$ and $H_{ijk}(X)=X^\ast H_{ijk}(x)$, with $x^i$ coordinates on $M$, denote pull-backs of $g$ and $H$ to $\Sigma_2$ and $\Sigma_3$, respectively. Even though, $H$ is defined on $\Sigma_3$, the Wess-Zumino term does not depend on the choice of $\Sigma_3$, or more precisely, it is ambiguous up to an integer constant, but the corresponding path integral is not\cite{Witten:1983ar}.

Given the action functional \eqref{NLSM}, it is possible to look for its extensions by additional 1-form gauge field $A$ such that the resulting action represents the gauging of the original one, and as such, is equal to the original one when all the gauge fields are set to zero. The usual way to do this is to take some Lie algebra $\mf{g}$ that can act on $M$ via Lie algebra homomorphism $\rho:\mf{g}\rightarrow\Gamma(TM)$. Then the vector fields in the image of $\rho$ create a foliation of $M$. In general, this foliation can be singular, meaning that the gauge orbits are not all of the same dimension. Actually, it is quite common to have singular foliations. 

Another way to view the action of $\mf{g}$ on $M$ is to notice that $M\times\mf{g}$ forms a Lie algebroid over $M$ with $\rho$ as an anchor. However, as described in \cite{ChatzistavrakidisAHP, Kotov:2014iha}, the gauging of the action \eqref{NLSM} can be considered for a much wider class of singular foliations. One way to consider more general gauging is to replace $M\times\mf{g}$ with a general Lie algebroid $E$. Being a Lie algebroid, $E$ is equipped with a Lie bracket $[\cdot,\cdot]_E:\Gamma(E)\times\Gamma(E)\rightarrow\Gamma(E)$ that satisfies the Jacobi identity:
\begin{equation}
[[e,e']_E,e'']_E+[[e',e'']_E,e]_E+[[e'',e]_E,e']_E=0\,,\qquad\forall e,e',e''\in\Gamma(E)\,.
\end{equation}
Let $e_a$ be a local basis for sections of $E$. Then for this basis it is possible to define structure functions $C$ such that:
\begin{equation}
[e_a,e_b]_E=\tensor{C}{^c_{ab}}e_c\,.
\end{equation}
Using the anchor $\rho$ of the Lie algebroid $E$, it also possible to define vector fields $\rho_a=\rho(e_a)=\rho_a^i\partial_i$. Since $\rho$ is the Lie algebra homomorphism, these vector fields have the same structure functions as the local basis sections $e_a$:
\begin{equation}
[\rho_a,\rho_b]=\tensor{C}{^c_{ab}}\rho_c\,,
\label{algebra_closure}
\end{equation}
and the Jacobi identity of the Lie bracket $[\cdot,\cdot]_E$ gives rise to an identity:
\begin{equation}
\iota_{\rho_{[a}}\dd \tensor{C}{^d_{bc]}}=\tensor{C}{^d_{e[a}}\tensor{C}{^e_{bc]}}\,.
\end{equation}

Given the Lie algebroid $E$ and the corresponding foliation on $M$, the question becomes whether there exist an action $S_0[X,A]$, with $A\in\Omega^1(\Sigma_2,X^\ast E)$ being 1-form gauge fields, such that $S_0$ reduces to \eqref{NLSM} when $A$ is set to 0. This action can be written quite generally as:
\begin{equation}
S_0[X,A]=S[X]-\int_{\Sigma_2}\left(A^a\wedge \theta_a(X) +A^a\wedge\ast\widetilde{\theta}_a(X)+\frac{1}{2}\gamma_{ab}(X)A^a\wedge A_b+\frac{1}{2}\widetilde{\gamma}_{ab}(X)A^a\wedge\ast A^b\right)\,,
\label{gauged_action}
\end{equation}
where $\theta_a(X)=\theta_{ai}(X)\dd X^i$ and $\widetilde{\theta}_a=\widetilde{\theta}_{ai}(X)\dd X^i$ are 1-forms, and $\gamma_{ab}$ and $\widetilde{\gamma}_{ab}$ functions on $M$, all pulled back to $\Sigma_2$ by $X$. This gauged action is still quite general and it is necessary to find conditions on the background fields $g$ and $H$, as well as the constraints on the gauging data $\theta$, $\widetilde{\theta}$, $\gamma$, $\widetilde{\gamma}$ and $\rho$ for this gauging to be possible. To do that we specify the following gauge transformations:
\begin{eqnarray}
\delta X^i &=& \rho^i_a(X)\epsilon_a\,,\label{gauge_trans_X}\\[4pt]
\delta A^a &=& \tensor{r}{^a_b}(X)\dd\epsilon^b+\tensor{s}{^a_b}(X)\ast\dd\epsilon^b+\tensor{C}{^a_{bc}}(X)A^b\epsilon^c+\tensor{\omega}{^a_{bi}}(X)\epsilon^b F^i+\tensor{\phi}{^a_{bi}}(X)\epsilon^b\ast F^i+\nn\\[4pt]
&&+\tensor{\chi}{^a_{bc}}(X)A^b\epsilon^c+\tensor{\psi}{^a_{bc}}(X)\ast A^b \epsilon^c\,,\label{gauge_trans_A}
\end{eqnarray}
where $\epsilon^a\in\Gamma(X^\ast E)$ is the scalar gauge parameter, $\tensor{r}{^a_b}(X)$, $\tensor{s}{^a_b}(X)$, $\tensor{\chi}{^a_{bc}}(X)$ and $\tensor{\psi}{^a_{bc}}(X)$ are functions and $\tensor{\omega}{^a_b}(X)$ and $\tensor{\phi}{^a_b}$ 1-forms on $M$, pulled back to $\Sigma_2$. Here $C$ has been explicitly written apart from $\chi$ for future convenience. Furthermore, 1-form $F^i$ is a covariant exterior derivative of $X^i$ defined as:
\begin{equation}
F^i=\dd X^i-\rho^i_a A^a\,.
\end{equation}
Given these gauge transformations of $X$ and $A$, it is easy to find the gauge transformation of the action \eqref{gauged_action}. Requirement that such a transformation vanish gives conditions for the fields $g$ and $H$:
\begin{eqnarray}
\mathcal{L}_{\rho_a}g&=&-\tensor{\omega}{^b_a}\vee\widetilde{\theta}_b+\tensor{\phi}{^b_a}\vee\theta_b\,,\\[4pt]
\iota_{\rho_a}H &=& \dd\left(\tensor{r}{^b_a}\theta_b-\tensor{s}{^b_a}\widetilde{\theta}_b\right)-\tensor{\omega}{^b_a}\wedge\theta_b+\tensor{\phi}{^b_a}\wedge\widetilde{\theta}_b\,,\\[4pt]
\iota_{\rho_a}g &=& \tensor{s}{^b_a}\theta_b-\tensor{r}{^b_a}\widetilde{\theta}_b\,,
\end{eqnarray}
in addition to the following constraints:
\begin{eqnarray}
\iota_{\rho_a}\theta_b &=& \tensor{r}{^c_a}\gamma_{cb}-\tensor{s}{^c_a}\widetilde{\gamma}_{cb} \,,\\[4pt]
\iota_{\rho_a}\widetilde{\theta}_b &=& \tensor{s}{^c_a}\gamma_{cb}-\tensor{r}{^c_a}\widetilde{\gamma}_{cb} \,,\\[4pt]
\mathcal{L}_{\rho_a}\theta_b &=& -\tensor{(C+\chi)}{^c_{ba}}\theta_c+\tensor{\psi}{^c_{ba}}\widetilde{\theta}_c+\iota_{\rho_b}\dd\left(\tensor{r}{^c_a}\theta_c-\tensor{s}{^c_a}\widetilde{\theta}_c\right)+\iota_{\rho_a}\iota_{\rho_b}H+\nn\\[4pt]
&&+\left(\gamma_{cb}+\iota_{\rho_b}\theta_c\right)\tensor{\omega}{^c_a}-\left(\widetilde{\gamma}_{cb}+\iota_{\rho_b}\widetilde{\theta}_c\right)\tensor{\phi}{^c_a}\,,\\[4pt]
\mathcal{L}_{\rho_a}\widetilde{\theta}_b &=& -\tensor{(C+\chi)}{^c_{ba}}\widetilde{\theta}_c+\tensor{\psi}{^c_{ba}}\theta_c-\iota_{\rho_b}\mathcal{L}_{\rho_a}g+\nn\\[4pt]
&&+\left(\gamma_{cb}+\iota_{\rho_b}\theta_c\right)\tensor{\phi}{^c_a}-\left(\widetilde{\gamma}_{cb}+\iota_{\rho_b}\widetilde{\theta}_c\right)\tensor{\omega}{^c_a}\,,\\[4pt]
\frac{1}{2}\mathcal{L}_{\rho_a}\gamma_{cb} &=& \gamma_{d[c}\tensor{(C+\chi)}{^d_{b]a}}-\widetilde{\gamma}_{d[c}\tensor{\psi}{^d_{b]a}}-\gamma_{d[c}\iota_{\rho_{b]}}\tensor{\omega}{^d_a}+\widetilde{\gamma}_{d[c}\iota_{\rho_{b]}}\tensor{\phi}{^d_a}\,,\\[4pt]
\frac{1}{2}\mathcal{L}_{\rho_a}\widetilde{\gamma}_{cb} &=& -\widetilde{\gamma}_{d[c}\tensor{(C+\chi)}{^d_{b]a}}+\gamma_{d[c}\tensor{\psi}{^d_{b]a}}+\widetilde{\gamma}_{d[c}\iota_{\rho_{b]}}\tensor{\omega}{^d_a}-\gamma_{d[c}\iota_{\rho_{b]}}\tensor{\phi}{^d_a}\,.
\end{eqnarray}

At this point it is convenient to consider redefined quantities:
\begin{eqnarray}
\theta^\pm_a &=&\theta_a\pm\widetilde{\theta}_a\,,\\[4pt]
\gamma^\pm_{ab} &=& \gamma_{ab}\pm\widetilde{\gamma}_{ab}\,,\\[4pt]
\tensor{r}{^{\pm a}_b} &=& \frac{1}{2}\left(\tensor{r}{^a_b}\pm\tensor{s}{^a_b}\right)\,,\\[4pt]
\tensor{\Omega}{^{\pm a}_b} &=& \tensor{\omega}{^a_b}\pm\tensor{\phi}{^a_b}\,,\\[4pt]
\tensor{\mathcal{C}}{^{\pm c}_{ab}} &=& -\tensor{C}{^c_{ab}}-\tensor{\chi}{^c_{ab}}\mp\tensor{\psi}{^c_{ab}}+\frac{1}{2}\iota_{\rho_a}\tensor{\Omega}{^{\pm c}_b}\,.
\end{eqnarray}
In the terms of these new quantities, the above conditions become:
\begin{eqnarray}
\mathcal{L}_{\rho_a}g &=& \frac{1}{2}\left(\tensor{\Omega}{^{+b}_a}\vee\theta^-_b-\tensor{\Omega}{^{-b}_a}\vee\theta^+_b\right)\,,\label{condition1}\\[4pt]
\iota_{\rho_a}H &=& \dd\left(\tensor{r}{^{+b}_a}\theta^-_b+\tensor{r}{^{-b}_a}\theta^+_b\right)-\frac{1}{2}\left(\tensor{\Omega}{^{+b}_a}\wedge\theta^-_b+\tensor{\Omega}{^{-b}_a}\wedge\theta^+_b\right)\,,\label{condition2}\\[4pt]
\iota_{\rho_a}g &=& \tensor{r}{^{+b}_a}\theta^-_b-\tensor{r}{^{-b}_a}\theta^+_b\,,\label{condition3}
\end{eqnarray}
and the constraints simplify to:
\begin{eqnarray}
\frac{1}{2}\iota_{\rho_a}\theta^\pm_b &=& \tensor{r}{^{\pm c}_a}\gamma^\mp_{cb}\,,\label{constraint1}\\[4pt]
\mathcal{L}_{\rho_a}\theta^\pm_b &=& \tensor{\mathcal{C}}{^{\mp c}_{ba}}\theta^\pm_c+\frac{1}{2}\tensor{\Omega}{^{\pm c}_a}\gamma^\mp_{cb}\,,\label{constraint2}\\[4pt]
\frac{1}{2}\mathcal{L}_{\rho_a}\gamma^\pm_{cb} &=& \gamma^\mp_{d[b}\tensor{\mathcal{C}}{^{\pm d}_{c]a}}\,.\label{constraint3}
\end{eqnarray}

Under the assumption that $r^\pm$ are invertible, it is possible to redefine the gauge field as:
\begin{equation}
\widetilde{A}^a=\frac{1}{4}\tensor{\left((r^+)^{-1}\right)}{^a_b}(A^b+\ast A^b)+\frac{1}{4}\tensor{\left((r^-)^{-1}\right)}{^a_b}(A^b-\ast A^b)\,.
\end{equation}
With this new field, using \eqref{condition3} and \eqref{constraint1}, the gauged action becomes:
\begin{equation}
S_0[X,\widetilde{A}]=-\int_{\Sigma_2}\left(\frac{1}{2}g_{ij}(X)\widetilde{F}^i\wedge\ast 	\widetilde{F}^j+\widetilde{A}^a\wedge\theta'_a(X)+\frac{1}{2}\gamma'_{ab}(X)\widetilde{A}^a\wedge\widetilde{A}^b\right)-\int_{\Sigma_3}H(X)\,,
\end{equation}
where $\widetilde{F}$, $\theta'_a$ and $\gamma'_{ab}$ are defined as:
\begin{eqnarray}
\widetilde{F}^i &=& \dd X^i-\rho^i_a \widetilde{A}^a\,,\\[4pt]
\theta'_a &=& \tensor{r}{^{-b}_a}\theta^+_b+\tensor{r}{^{+b}_a}\theta^-_b\,,\\[4pt]
\gamma'_{ab} &=&2\tensor{r}{^{-c}_a}\tensor{r}{^{+d}_b}\gamma^+_{cd}+2\tensor{r}{^{+c}_a}\tensor{r}{^{-d}_b}\gamma^-_{cd}\,.
\end{eqnarray}
But this is just the minimal coupling to the metric sector. So the only nonstandard gauging is in the topological sector. This also forces the simplification of gauge transformations to:
\begin{equation}
\delta\widetilde{A}^a=\dd\epsilon^a+\tensor{C}{^a_{bc}}(X)\widetilde{A}^b\epsilon^c+\tensor{{\omega'}}{^{a}_{bi}}(X)\epsilon^b \widetilde{F}^i+\tensor{{\phi'}}{^a_{bi}}(X)\epsilon^b\ast \widetilde{F}^i\,,
\end{equation}
with $\tensor{{\omega'}}{^{a}_{b}}$ and $\tensor{{\phi'}}{^a_{b}}$ defined as:
\begin{eqnarray}
\tensor{{\omega'}}{^{a}_{b}} &=& \frac{1}{4}\tensor{\left((r^+)^{-1}\right)}{^a_c}\tensor{\Omega}{^{+c}_b}+\frac{1}{4}\tensor{\left((r^-)^{-1}\right)}{^a_c}\tensor{\Omega}{^{-c}_b}\,,\\[4pt]
\tensor{{\phi'}}{^{a}_{b}} &=& \frac{1}{4}\tensor{\left((r^+)^{-1}\right)}{^a_c}\tensor{\Omega}{^{+c}_b}-\frac{1}{4}\tensor{\left((r^-)^{-1}\right)}{^a_c}\tensor{\Omega}{^{-c}_b}\,.
\end{eqnarray}

The above simplification looks valid only if $r^\pm$ are invertible. However, one can come to the same conclusion for a much wider class of possibilities. For the sake of simplicity, we shall assume $r^-$ not to be invertible. In order to simplify this situation, one can always change the frame in order to write $r^-$ in block-diagonal form such that one block is nilpotent (such that it has only 0 eigenvalues), and the other is invertible. The only one of interest here is the nilpotent part so one can assume for the whole $r^-$ to be nilpotent, as the general situation is then obtained by combining the reversible and nilpotent cases. If this nilpotent part is actually 0, plugging this into the gauging conditions  and constraints forces $\iota_{\rho_a}g$ to vanish. Since the metric cannot have zero eigenvalues, this implies that all $\rho_a$ vanish, thus leaving us with no gauging at all. So, one concludes that while it is possible to gauge with noninvertible $r^\pm$, such gauging leads to redundant degrees of freedom in the gauge field that can always be removed, as long as the nilpotent part of $r^\pm$ is semisimple as well. Thus, the gauging in the metric sector can always be put in the form of minimal coupling.

In what follows, we shall only consider this simplified version of the gauged action, assuming that all the necessary redefinitions have been made.

\subsection{Dirac sigma models as gauge theory}

Given the simplified gauged action:
\begin{equation}
S_0[X,A]=-\int_{\Sigma_2}\left(\frac{1}{2}g_{ij}(X)F^i\wedge\ast F^j+A^a\wedge\theta_a(X)+\frac{1}{2}\gamma_{ab}(X)\widetilde{A}^a\wedge\widetilde{A}^b\right)-\int_{\Sigma_3}H(X)\,,
\label{DSM}
\end{equation}
and the corresponding gauge transformations:
\begin{eqnarray}
\delta X^i &=& \rho^i_a\epsilon^a\,,\label{gaugeX}\\[4pt]
\delta A^a &=& \dd\epsilon^a +\tensor{C}{^a_{bc}}A^b\epsilon^c+\tensor{\omega}{^a_{bi}}\epsilon^b F^i+\tensor{\phi}{^a_{bi}}\epsilon^b\ast F^i\,,\label{gaugeA}
\end{eqnarray}
there still remains the question of what the gauging conditions and constraints \eqref{condition1}-\eqref{constraint3} tell us about gauging data. This has been explored in detail in \cite{ChatzistavrakidisAHP}. First, the conditions on the background fields are:
\begin{eqnarray}
\mathcal{L}_{\rho_a}g &=& \tensor{\omega}{^b_a}\vee\iota_{\rho_b}g+\tensor{\phi}{^b_a}\vee\theta_b\,,\label{con1}\\[4pt]
\iota_{\rho_a}H &=& \dd\theta_a-\tensor{\omega}{^b_a}\wedge\theta_b-\tensor{\phi}{^b_a}\wedge\iota_{\rho_b}g\,.\label{con2}
\end{eqnarray}
In addition there are 2 additional constraints. The first of those specifies $\gamma$:
\begin{equation}
\gamma_{ab}=\iota_{\rho_a}\theta_b\,,
\end{equation}
and since $\gamma_{ab}$ is antisymmetric, this gives another constraint on $\theta$:
\begin{equation}
\iota_{\rho_a}\theta_b+\iota_{\rho_b}\theta_a=0\,.
\label{invol}
\end{equation}
The final constraint is:
\begin{equation}
\tensor{C}{^c_{ab}}\theta_c=\mathcal{L}_{\rho_a}\theta_b-\iota_{\rho_b}\dd\theta_a-\iota_{\rho_a}\iota_{\rho_b}H\,.
\label{Courant_closure}
\end{equation}
These last two conditions can be combined with the closure of the Lie algebra \eqref{algebra_closure} of the vector fields $\rho_a$ to give an interesting geometric interpretation of the constraints. In the generalized tangent bundle $TM\oplus T^\ast M$, viewed as an $H$-twisted Courant algebroid, sections $\rho_a+\theta_a$ live in a specific subbundle, called the Dirac structure, that is isotropic with the respect to the pairing, and closed under the action of the twisted Courant bracket. If these Dirac structures are of maximal rank, then the corresponding gauge theory is topological, and otherwise it is nontopological. In this paper we consider only Dirac structures of maximal rank.

As for the constraints, it is desirable to find the geometric (frame independent) form of the conditions \eqref{con1} and \eqref{con2}. The first thing to do here is to look at $\omega$ and $\phi$, defined as coefficients in the gauge transformation of the gauge field $A$. By computing the transformations of these coefficients under the frame change, it turns out that $\phi$ transforms tensorially but $\omega$ does not, but instead transforms as a connection. As such $\tensor{\omega}{^a_{bi}}$ can be interpreted as the components of a connection $\nabla^\omega:\Gamma(E)\rightarrow\Gamma(T^\ast M\otimes E)$ on $E$ such that:
\begin{equation}
\nabla^\omega e_a=\tensor{\omega}{^a_b}\otimes e_b\,.
\end{equation}
Furthermore, since $\phi$ transforms tensorially, it can be interpreted as an endomorphism on $E$. As a result $\Omega^\pm$ both transform as the components of the connections $\nabla^\pm$ on $E$:
\begin{equation}
\nabla^\pm e_a=\tensor{\Omega}{^{\pm b}_a}\otimes e_b\,.
\end{equation}
In the following, we shall use $\Omega^\pm$ instead of $\omega$ and $\phi$ since it turns out to be more convenient.

Finally, to express conditions \eqref{con1} and \eqref{con2} in frame independent form, we look at sections:
\begin{equation}
\mathcal{G}_\pm=\theta\pm\rho^\ast\in\Gamma(T^\ast M\otimes E\ast)\,,
\end{equation}
where $\rho^\ast=\iota_{\rho_a}g\otimes e^a$. When written in terms of these sections, the conditions \eqref{con1} and \eqref{con2} take the form:
\begin{eqnarray}
&&\text{Sym}\left(\nabla^+\mathcal{G}_+-\nabla^-\mathcal{G}_-\right)=0\,,\label{fcon1}\\[4pt]
&&D^+\mathcal{G}_+ +D^-\mathcal{G}_-=2\iota_\rho H\,,\label{fcon2}
\end{eqnarray}
where $D^\pm$ are the exterior derivatives associated to $\nabla^\pm$.

\subsection{Field equations, curvature and torsion}

For the topological Dirac sigma models, the corresponding Dirac structure is of maximal rank, or in other words, its rank is equal to the dimension of $M$. This means that, as have been shown in \cite{Chatzistavrakidis:2016jci}, even though $\rho$ and $\theta$ aren't invertible by themselves, their combined sections $\mathcal{G}_\pm$ are. As a result, it is possible to write explicit expressions for $\Omega^\pm$ by inverting \eqref{fcon1} and \eqref{fcon2}:
\begin{equation}
\tensor{\Omega}{^{\pm a}_{bi}}=(\mathcal{G}_\pm^{-1})^{aj}\left(\partial_i\mathcal{G}_{\pm bj}-\mathring{\Gamma}^{k}_{ji}\mathcal{G}_{\pm bk}-\frac{1}{2}\rho_b^k H_{ijk}\right)\,,
\end{equation}
where $\mathring{\Gamma}^k_{ij}$ are the coefficients of the Levi-Civita connection on $M$.

The invertibility of $\mathcal{G}_\pm$ has an interesting consequence on the field equations. A functional derivative of the action \eqref{DSM} with respect to $A$ produces the equation:
\begin{equation}
\left(\theta_{ai}-(\iota_{\rho_a}g)_i\ast\right)F^i=0\,.
\end{equation}
For Dirac structures, the operator in the brackets is invertible, which then simplifies the field equation to:
\begin{equation}
F^i=0\,.
\end{equation}
The other field equation is obtained through a functional derivative of the action \eqref{DSM} with respect to $X$:
\begin{equation}
G_i=\dd (\theta_{ai} A^a)+\frac{1}{2}\left(\rho_b^j\partial_i\theta_{aj}-\theta_{aj}\partial_i\rho_b^j+(\iota_{\rho_b}\iota_{\rho_a}H)_i\right)A^a\wedge A^b=0\,.
\end{equation}

The invertibility of $\mathcal{G}_\pm$ can be used to induce connections $\nabla^{\ast\pm}$ on $T^\ast M$ by the connections $\nabla^\pm$ on $E$ through:
\begin{equation}
\nabla^{\ast\pm}=\mathcal{G}_\pm\circ\nabla^\pm\circ\mathcal{G}_\pm^{-1}\,.
\end{equation}
Then the coefficients of these induced connections are:
\begin{equation}
{\Gamma^{\ast\pm}}^k_{ij}=-\mathring{\Gamma}^k_{ij}+\frac{1}{2}(\mathcal{G}_\pm^{-1})^{ka}(\iota_{\rho_a}H)_{ij}\,.
\end{equation}
But these connections on $T^\ast M$ then induce connections $\nabla^{\sharp\pm}$ on $TM$ as dual connections of $\nabla^{\ast\pm}$. Its coefficients are then:
\begin{equation}
{\Gamma^\pm}^k_{ij}=\mathring{\Gamma}^k_{ij}-\frac{1}{2}(\mathcal{G}_\pm^{-1})^{ka}(\iota_{\rho_a}H)_{ij}\,.
\end{equation}
The torsion tensor of these connections is equal to:
\begin{equation}
\tensor{\Theta}{^{\pm k}_{ij}}=-2{\Gamma^\pm}^k_{[ij]}=(\mathcal{G}_\pm^{-})^{ka}(\iota_{\rho_a}H)_{ij}\,.
\end{equation}
So, the torsion of induced connections on $TM$ is controlled by the 3-form $H$.

Other then induced connections on $TM$, one can define the curvature of connections $\nabla^\pm$ on $E$ in a standard way:
\begin{equation}
\tensor{R}{^{\pm a}_b}=\dd\tensor{\Omega}{^{\pm a}_b}+\tensor{\Omega}{^{\pm a}_c}\wedge\tensor{\Omega}{^{\pm c}_b}\,.
\end{equation}
This curvature satisfies the Bianchi identity:
\begin{equation}
\nabla_{[i}^\pm\tensor{R}{^{\pm a}_{bjk]}}+\tensor{\Theta}{^{\pm l}_{[ij}}\tensor{R}{^{\pm a}_{bk]l}}=0\,.
\label{bianchi_theta}
\end{equation}

Other then standard curvature and torsion, there is a notion of $E$-curvature and $E$-torsion, related to $E$-connection $^E\nabla$ and $E$-covariant derivative. An ordinary covariant derivative $\nabla_v$ is defined along a section $v\in\Gamma(TM)$ from the tangent bundle. An $E$-covariant derivative is a generalization in which a derivative is not defined necessarily along a section from the tangent bundle, but along a section from the Lie algebroid $E$. Furthermore, the ordinary covariant derivative $\nabla_v$, for a fixed $v\in\Gamma(TM)$, is a linear function $\nabla_v:\Gamma(TM)\rightarrow\Gamma(TM)$. For $E$-covariant derivative this is generalized such that instead of $TM$, an arbitrary bundle $\hat{E}$ is used. As such, $^E\nabla_e:\Gamma(\hat{E})\rightarrow\Gamma(\hat{E})$, for $e\in\Gamma(E)$. Just as an ordinary derivative, it is linear and it satisfies the Leibniz rule:
\begin{equation}
^E\nabla_e(f\hat{e})=f\, ^E\nabla_e\hat{e}+(\rho(e)f)\hat{e}\,,\quad\forall e\in\Gamma(E)\,,\forall\hat{e}\in\Gamma(\hat{E})\,.
\end{equation}
This then also introduces a notion of an $E$ curvature:
\begin{equation}
^E R(e,e')=[^E\nabla_e,^E\nabla_{e'}]-^E\nabla_{[e,e']}\,.
\end{equation}
Furthermore, if $\hat{E}=E$, then a notion of an $E$-torsion can be introduced as well:
\begin{equation}
^E T(e,e')=^E\nabla_e e'-^E\nabla_{e'}e-[e,e']\,.
\end{equation}
Finally, it is often convenient to define a quantity:
\begin{equation}
S=\nabla(^E T)+2\text{Alt}(\iota_\rho R)\,,
\end{equation}
which is called the basic curvature in \cite{Ikeda:2019czt}.

In the present situation, the two ordinary curvatures $\nabla^\pm$ on $E$ induce two $E$-connections on $E$ through a simple identification:
\begin{equation}
^E\nabla^\pm_e e'=\nabla^\pm_{\rho(e)} e'\,.
\end{equation}
Then the corresponding $E$-torsion equals:
\begin{equation}
\tensor{T}{^{\pm c}_{ab}}=-\tensor{C}{^c_{ab}}+2\iota_{\rho_{[a}}\tensor{\Omega}{^{\pm c}_b}\,,
\end{equation}
where we are using $T^\pm$ instead of $^E T^\pm$ to denote $E$-torsion. In addition to $E$-curvatures, there are two basic curvatures:
\begin{equation}
\tensor{S}{^{\pm c}_{ab}}=\nabla^\pm \tensor{T}{^{\pm c}_{ab}}+2\iota_{\rho_{[a}}\tensor{R}{^{\pm c}_{b]}}\,.
\end{equation}
These torsions and basic curvatures satisfy several identities that will be used when determining the BV action of the Dirac sigma model:
\begin{eqnarray}
\tensor{T}{^{\pm c}_{ab}}\rho_c^i &=& -2\rho_{[a}^j\nabla^\pm_j \rho_{b]}^i+\rho_a^j\rho_b^k\tensor{\Theta}{^{\pm i}_{jk}}\,,\label{eq1}\\
\iota_{\rho_{[b}}\iota_{\rho_a}\tensor{R}{^{\pm d}_{c]}} &=& \rho_{[a}^i \nabla^\pm_i\tensor{T}{^{\pm d}_{bc}}+\tensor{T}{^{\pm d}_{e[a}}\tensor{T}{^{\pm e}_{bc]}}\,,\label{eq2}\\
\left[\nabla^{\pm}_i,\nabla^{\pm}_j\right]\tensor{T}{^{\pm a}_{bc}} &=& -\tensor{\Theta}{^{\pm k}_{ij}}\nabla^\pm_k\tensor{T}{^{\pm a}_{bc}}+\tensor{T}{^{\pm d}_{bc}}\tensor{R}{^{\pm a}_{dij}}-\tensor{T}{^{\pm a}_{dc}}\tensor{R}{^{\pm d}_{bij}}-\tensor{T}{^{\pm a}_{bd}}\tensor{R}{^{\pm d}_{cij}}\,,\,\,\,\,\,\,\,\,\,\label{eq3}\\
\tensor{T}{^{\pm c}_{ab}} &=& (\mathcal{G}_\pm^{-1})^{ci}\left(\rho_{[a}^j\nabla^\pm_i\theta_{b]j}-\theta_{[bj}\nabla^\pm_i\rho_{a]}^j-\rho_{[a}^j\theta_{b]k}\tensor{\Theta}{^{\pm k}_{ji}}\right)\,,\label{eq4}
\end{eqnarray}
with $\nabla^\pm$ in these equations acts both as $\nabla^\pm$ when acting on bundle indices, and as $\nabla^{\sharp\pm}$ when acting on tangent indices.

\subsection{Target space covariance}

Up to now, the action, the field equations and the gauge transformations of the Dirac sigma model has been presented with manifest spacetime covariance, but not target space covariance. Here we show how the connections $\nabla^\pm$ guarantee this covariance.

Let us first consider gauge transformations \eqref{gaugeX} and \eqref{gaugeA}. The transformation of $X$ can be easily written in the basis-independent way, through the use of the anchor $\rho$:
\begin{equation}
\delta X=\rho(\epsilon)\,.
\end{equation}
For the other gauge transformation, one should first note that the 1-form gauge field is $A=A^a\otimes e_a$, meaning that the full transformation of $A$ should include, besides gauge transformation of $A^a$, the transformation coming from the frame change due to the change of base points. Any of the connections $\nabla^\pm$ can be used to take into account this change of frame so $\delta e_a=\tensor{\Omega}{^{\pm b}_{ai}}\delta X^i e_b$, which then gives the transformation of $A$:
\begin{equation}
\delta A=(\delta A^a+\iota_{\rho_c}\tensor{\Omega}{^{\pm a}_b}A^b\epsilon_c)\otimes e_a\,.
\end{equation}
Also, $\delta A^a$ has to be rewritten in terms of connection which then gives:
\begin{equation}
\delta A^a=\DD^\pm\epsilon^a-\left(\tensor{T}{^{\pm a}_{bc}}- \iota_{\rho_c}\tensor{\Omega}{^{\pm a}_b}\right)A^b\epsilon^c+\frac{1}{2}\left(\tensor{\Omega}{^{+a}_{bi}}(1+\ast)+\tensor{\Omega}{^{-a}_{bi}}(1-\ast)-\tensor{\Omega}{^{\pm a}_{bi}}\right)\epsilon^b F^i\,.
\end{equation}
This then expresses $\delta A$ in terms of $\Omega^+$ or $\Omega^-$. By adding those two options together, the final form of the gauge transformation is obtained:
\begin{equation}
\d A=\frac{\DD^{+}+\DD^{-}}{2}\,\epsilon-\frac {T^{+}+T^{-}}{2}\,(A,\epsilon)+\biggl\langle\frac {\Omega^{+}-\Omega^{-}}{2},\ast F\biggl\rangle(\epsilon)\,.
\end{equation}
Notice that $(\Omega^+-\Omega^-)/2$ is equal to the tensor $\phi$, so this is indeed the tensorial form of the gauge transformation.

Having written the gauge transformation in the target space covariant form, one can do the same with the field equations. The field equations for $A$ is already target space covariant:
\begin{equation}
F=\dd X-\rho(A)\,.
\end{equation}
To covariantize the other field equations, it is useful to define $\mf(a)=\theta(A)$. Then the field equation becomes:
\begin{equation}
G=\DD^{\pm}\mf{a}-\frac 12 \, T^{\pm}_{{\cal G}_{\pm}}(A,A)\mp \frac 12 \, \Theta_{\ast}^{\pm}(\rho(A),\rho(A),\cdot)\,,
\end{equation}
where $T^\pm_{\mathcal{G}_\pm}=\mathcal{G}_{\pm a}\otimes T^{\pm a}\in\Gamma(T^\ast M\otimes\Lambda^2 E^\ast)$ and $\Theta^\pm_\ast$ is the contraction of the torsion tensor $\Theta$ with the metric, such that its components are $\Theta^\pm_{\ast ijk}=g_{il}\tensor{\Theta}{^{\pm l}_{jk}}$.

Finally, the action \eqref{DSM} has to be rewritten in the manifestly target space covariant form. This is easy to do using the maps $\rho:E\rightarrow TM$ and $\theta:E\rightarrow T^\ast M$:
\begin{equation}
S_0=-\int_{\Sigma_2}\left(||F||^2+\left\langle (X^\ast\theta)(A),\dd X+\frac{1}{2}(X^\ast\rho)(A)\right\rangle\right)-\int_{\Sigma_3} X^\ast H\,,
\end{equation}
where $||F||^2=(X^\ast\rho)(F \overset{\wedge},\ast F)$.

\section{The BV action}

\subsection{BRST operator and antifields}

To find the BV action for the Dirac sigma models, we follow the standard steps described in \cite{HT, Gomis:1994he}, in a similar fashion as have been done in \cite{Ikeda:2019czt} for the twisted Poisson sigma model. First, the space of fields is enlarged by ghosts, one for each gauge parameter, and assign them a ghost number 1. In general, one would have to introduce ghosts for ghosts as well, but since the Dirac sigma model is irreducible, that is not necessary here. Since there is only one gauge parameter $\epsilon^a$ in the gauge transformations of the Dirac sigma model, only one ghost field $c^a$ introduced here. It is assigned the ghost number 1, denoted as $\gh{c^a}=1$. Next, the BRST operator $s$ is defined through its action on the fields:
\begin{eqnarray}
sX^{i} &=&\rho^{i}_a c^a\,,\\[4pt]
sA^a &=&\dd c^a+\tensor{C}{^a_{bc}}A^bc^c+ \tensor{\omega}{^a_{bi}}c^bF^{i}+\tensor{\phi}{^a_{bi}}c^b\ast F^{i}\,,\\[4pt]
sc^a &=&-\frac{1}{2}\tensor{C}{^a_{bc}}c^bc^c\,.
\end{eqnarray} 
The first two of these are just gauge transformation in which the gauge parameter $\epsilon^a$ is replaced by its corresponding ghost field $c^a$, while the action of $s$ on $c^a$ is defined such that the BRST operator is nilpotent on-shell.\footnote{The requirement that $s^2$ vanishes on-shell does not always uniquely determine the action of $s$ on ghost fields, but all the possible ambiguities are eventually taken into account through the construction of the BV action.} Using the commutation relations of the vectors $\rho_a$, and the Jacobi identity of the Lie algebroid for $\tensor{C}{^a_{bc}}$, it is straightforward to check that $s^2 X^i=0$ and $s^2 c^a =0$. However $s^2 A^a$ does not vanish identically, but is instead controlled by the field equations:
\begin{equation}
s^2A^a = \frac{1}{2}\tensor{S}{^a_{bci}}c^bc^cF^{i}+\frac{1}{2}\tensor{\widetilde{S}}{^a_{bci}}c^bc^c\ast F^{i}\,,
\end{equation}
where the curvature tensors $S$ and $\widetilde{S}$ are given by:
\begin{eqnarray}
\tensor{S}{^a_{bc}} &=& \frac{1}{2}\left(\tensor{S}{^{+a}_{bc}}+\tensor{S}{^{-a}_{bc}}\right)\,,\\
\tensor{\widetilde{S}}{^a_{bc}} &=& \frac{1}{2}\left(\tensor{S}{^{+a}_{bc}}-\tensor{S}{^{-a}_{bc}}\right)\,.
\end{eqnarray}
Since all the terms in $s^2 A^a$ contain field equations, it indeed does vanish on-shell, but not off-shell. This is due to the openness of the gauge algebra, meaning that it closes only on-shell. This implies that the BRST operator is not sufficient for the quantization of the theory, but a BV approach is necessary instead.

Having the BRST operator, the next step is to enlarge the space of fields by antifields for each of the fields. Having 3 fields already, $X^i$, $A^a$ and $c^a$, 3 antifields are necessary, $X^+_i$, $A^+_a$ and $c^+_a$. The form degree of these antifields is complementary to its corresponding field, while its ghost degree in a sum with the ghost degree of the corresponding field gives -1:
\begin{eqnarray}
\text{fdeg}(\Phi^+) &=& 2-\text{fdeg}(\Phi)\,,\\
\gh(\Phi^+) &=& -1-\gh(\Phi)\,,
\end{eqnarray}
where $\Phi$ denotes any of the fields. The ghost and form degrees of all the fields/antifields are collected in Table \ref{table1}.

	\begin{table}
\begin{center}	\begin{tabular}{| c | c | c | c | c | c | c |}
		\hline \multirow{3}{5em}{(Anti)Field} &&&&&& \\  & $X^{\m}$ & $A^a$ & $c^a$ & $X^{+}_{\m}$ & $A^{+}_a$ & $c^{+}_a$ \\ &&&&&& \\\hhline{|=|=|=|=|=|=|=|}
		\multirow{3}{4em}{Ghost degree} &&&&&& \\  & 0 & 0 & 1 & -1 & -1 & -2 \\ &&&&&& \\\hline 
		\multirow{3}{4em}{Form degree} &&&&&& \\  & 0 & 1 & 0 & 2 & 1 & 2 \\ &&&&&&
		\\\hline 
	\end{tabular}\end{center}\caption{The classical basis with ghost and form degrees for Dirac sigma models. }\label{table1}\end{table}
	
Finally, to be able to construct the BV action, a notion of antibracket is necessary. This can be introduced by defining a symplectic form on the space of fields:
\begin{equation}
\omega_{\text{BV}}=\int_\Sigma \left( \delta X^i\wedge \delta X^+_i +\delta A^a\wedge\delta A^+_a+\delta c^a\wedge \delta c^+_a\right)\,,
\label{sym_form}
\end{equation}
which induces the antibracket:
\begin{equation}
(F,G)_{\text{BV}}= \int \dd^2\sigma\, \dd^2\sigma' \sum_\Phi\left(\frac{\delta_R F}{\delta\Phi(\sigma)}\frac{\delta_L G}{\delta\Phi^\ast(\sigma')}-\frac{\delta_R F}{\delta\Phi^\ast(\sigma)}\frac{\delta_L G}{\delta\Phi(\sigma')}\right)\delta(\sigma-\sigma')\,,
\end{equation}
in term of the left and right functional derivative. Here, $\Phi^\ast$ and $\Phi^+$ are related through the Hodge dual operation, such that $\Phi^+=\ast\Phi^\ast$, while the left and right functional derivatives are defined through the variation of the action as:
\begin{equation}
\delta S =\int \sum_\Phi \delta\Phi\frac{\delta_L S}{\delta\Phi}=\int\sum_\Phi \frac{\delta_R S}{\delta\Phi} \delta\Phi\,.
\end{equation}

\subsection{Classical master equation}

The BV action $S_{BV}$ has to satisfy the classical master equation:
\begin{equation}
(S_{BV},S_{BV})=0\,.
\label{CME}
\end{equation}
In order to find $S_{BV}$ we can expand in terms of the numbers of antifields:
\begin{equation}
S_{BV}=S_0+S_1+S_2+\ldots\,,
\label{expansion}
\end{equation}
where the subscripts denote the number of antifields, such that $S_0$ does not contain antifields, $S_1$ contain terms with one antifield, and so on. The part without antifields here $S_0$ is just the classical action, while the part with one antifield $S_1$ is determined through the BRST transformations:
\begin{equation}
S_{1}=\int_{\Sigma_2}\left(X_{i}^{+}sX^{i}-A^{+}_a\w sA^{a}-c_a^{+}sc^{a}\right)\,.
\end{equation}
The part with two antifields $S_2$ has to be determined through the use of the classical master equation. In anticipation, of the result, we make the following ansatz:
\begin{equation}
S_{2}=-\int_\Sigma \frac 14\left( Y^{ab}_{cd}(X)\,A^{+}_a\w A^{+}_b+Z^{ab}_{cd}(X)\, A^{+}_{a}\w\ast A^{+}_{b} \right)c^cc^d\,,
\label{CME2}
\end{equation}
where $Y$ and $Z$ are $X$-dependent, for now undetermined functions. They are both antisymmetric in the lower indices, while $Y$ is symmetric and $Z$ antisymmetric in the upper indices.

Now the classical master equation \eqref{CME} has to be implemented. Using the expansion \eqref{expansion} in the number of antifields, the whole equation can be separated into smaller sectors. First, in the sector with no antifields there is one equation $(S_0,S_0)=0$ which is automatically satisfied because $S_0$ does not contain any antifields. Then, in the sector with one antifield, there is also one equation $(S_0,S_1)=0$. This is guaranteed by the on-shell nilpotency of the BRST operator, but it can also be easily checked by a straightforward calculation. The next step is the 2 antifield sector in which equation becomes:
\begin{equation}
(S_1,S_1)+2(S_0,S_2)=0\,.
\end{equation}
Calculation of each of these terms produces:
\begin{eqnarray}
\left(S_{1},S_{1}\right)&=& \int  \left(\tensor{\widetilde{S}}{^{a}_{cdi}}F^i\wedge\ast A^+_a-\tensor{S}{^{a}_{cdi}}F^i\wedge A^+_a\right)c^cc^d\,, \\[4pt]
\left(S_{0},S_{2}\right)&=&\int  \frac 12 \left[\left(Y^{(ab)}_{cd}\,(\iota_{\rho_b}g)_i-Z^{[ab]}_{cd}\,\theta_{bi}\right)F^i\wedge\ast A^+_a\right.+ \nn\\[4pt]  && \qquad\qquad \left. +\left(Y^{(ab)}_{cd}\,\theta_{bi}-Z^{[ab]}_{cd}\,(\iota_{\rho_b}g)_i\right)F^i\wedge A^+_a\right]c^cc^d\,.
\end{eqnarray}
Imposing \eqref{CME2} immediately gives $S$ and $\widetilde{S}$ in terms of $Y$ and $Z$:
\begin{eqnarray}
\tensor{S}{^{a}_{cd}}&=&Y^{(ab)}_{cd}\theta_b-Z^{[ab]}_{cd}\iota_{\r_{b}}g\,,
\\[4pt]
\tensor{\widetilde{S}}{^{a}_{cd}}&=&-Y^{(ab)}_{cd}\iota_{\r_{b}}g+Z^{[ab]}_{cd}\theta_b\,.
\end{eqnarray}
By defining:
\begin{equation}
Y^\pm=Y\pm Z\,,
\end{equation}
the above equations become:
\begin{equation}
\tensor{S}{^{\pm a}_{cd}}={Y^\pm}^{ab}_{cd}\mathcal{G}_{\mp b}\,.
\end{equation}
Since $\mathcal{G}_\pm$ are invertible, it is possible to write $Y^\pm$ in terms of $S^\pm$:
\begin{equation}
Y^{\pm}{}_{cd}^{ab}=\langle( {\cal G}_{\mp}^{-1})^{b},\tensor{S}{^{\pm a}_{cd}}\rangle\,.
\label{ypm}
\end{equation}

Finally, we have sectors with higher number of antifields, specifically, 3 and 4, which give equations:
\begin{eqnarray}
(S_1,S_2)+(S_0,S_3)&=&0\,\\
(S_2,S_2)+2(S_1,S_3)+2(S_0,S_4)&=&0\,.
\end{eqnarray}
It immediately evident that $(S_2,S_2)$ vanishes. This means that if $(S_1,S_2)$ vanishes as well, the classical master equation could be satisfied with $S_3, S_4, \ldots$ being equal to 0. So the next step is to evaluate $(S_1,S_2)$. A straightforward calculation leads to:
\begin{equation}
\left(S_{1},S_{2}\right)=- \int_{\Sigma_2}  \,\frac 14 \left( I^{ab}_{cde}A^+_{a}\wedge A^+_{b}+J^{ab}_{cde}A^+_{a}\wedge\ast A^+_{b}\right)c^cc^dc^{e}\,,
\end{equation}
where $I$ and $J$ are given by:
\begin{eqnarray}
I^{ab}_{cde}&=&\rho^{i}_{[e}\partial_{i}Y^{ab}_{cd]}-2(\tensor{C}{^{(a}_{p[e}}-\rho^{i}_{p}\tensor{\omega}{^{(a}_{[ei}})Y^{b)p}_{cd]}-2\rho^{i}_{p}\tensor{\phi}{^{(a}_{[e i}}Z^{b)p}_{cd]}-Y^{ab}_{p[e}\tensor{C}{^{p}_{cd]}}\,,
\\[4pt]
J^{ab}_{cde}&=&\rho^{i}_{[e}\partial_{i}Z^{ab}_{cd]}+2(\tensor{C}{^{[a}_{p[e}}-\rho^{i}_{p}\tensor{\omega}{^{[a}_{[ei}})Z^{b]p}_{cd]}+2\rho^{i}_{p}\tensor{\phi}{^{[a}_{[e i}}Y^{b]p}_{cd]}-Z^{ab}_{p[e}\tensor{C}{^{p}_{cd]}}\,.
\end{eqnarray}
By adding and subtracting, these equations can be expressed in terms of $Y^\pm$, and then, through \eqref{ypm}, in terms of $S_\pm$. The resulting equations turn out to be Bianchi identities for the basic curvatures $S$:
\begin{eqnarray}
&&\rho^{i}_{[e}\nabla^{\pm}_{i}\langle({\mc G}_{\mp}^{-1})^{b},\tensor{S}{^{\pm a}_{cd]}}\rangle-\tensor{T}{^{\pm f}_{[cd}}\langle ({\mc G}_{\mp}^{-1})^{b},\tensor{S}{^{\pm a}_{e]f}}\rangle+ \tensor{T}{^{\pm a}_{f[e}}\langle ({\mc G}_{\mp}^{-1})^{b},\tensor{S}{^{\pm f}_{cd]}}\rangle + \nn\\[4pt] && \hspace{225pt} + \tensor{T}{^{\mp b}_{f[e}}\langle ({\mc G}_{\mp}^{-1})^{f},\tensor{S}{^{\pm a}_{cd]}}\rangle=0\,,
\end{eqnarray}
which can be proven by a direct calculation using identities \eqref{bianchi_theta} and \eqref{eq1}-\eqref{eq4}. So the antibracket of $S_1$ and $S_2$ does indeed vanish meaning that the BV action contains terms that are at most quadratic in antifields. The whole BV action is then:
\begin{eqnarray}
S_{BV}&=&-\int_{\S_2} \left(\frac 12 g_{ij}(X)  F^{i}\w\ast F^{j}+A^a\w\theta_a(X)+\frac 12 \g_{ab}(X)A^a\w A^b\right)-\int_{\widehat{\Sigma}}X^{\ast}H \nn\\[4pt] 
&& +\int_{\S_2}\left(\r_{a}^{i}(X)X_{i}^{+}c^{a}+\frac 12\tensor{C}{^{a}_{bc}}(X) c_a^{+}c^{b}c^{c}\right) \nn\\[4pt] 
&& -\int_{\S_2} A^{+}_a\w \left(\dd c^{a}+\tensor{C}{^{a}_{bc}}(X)A^{b}c^{c}+\tensor{\omega}{^{a}_{bi}}(X)c^{b} F^{i}+\tensor{\phi}{^{a}_{bi}}(X)c^{b}\ast F^{i}\right) \nn\\[4pt] 
&&- \int_{\S_2} \frac 18 \left(\langle( {\cal G}_{-}^{-1})^{b},\tensor{S}{^{+a}_{cd}}\rangle(X)+\langle( {\cal G}_{+}^{-1})^{b},\tensor{S}{^{-a}_{cd}}\rangle(X)\right)\,A^{+}_a\w A^{+}_b c^cc^d
\nn\\[4pt]
&&-\int_{\S_2}\frac 18 \left(\langle( {\cal G}_{-}^{-1})^{b},\tensor{S}{^{+a}_{cd}}\rangle(X)-\langle( {\cal G}_{+}^{-1})^{b},\tensor{S}{^{-a}_{cd}}\rangle(X)\right)\, A^{+}_{a}\w\ast A^{+}_{b} c^cc^d\,.
\label{bv}
\end{eqnarray}

\subsection{Manifestly target space covariant BV action}

Since the classical action \eqref{DSM} is target space covariant, one expects the same to be true for the BV action as well. This covariance is not manifest however, mainly due to the term involving $X^+$.

First thing to notice is that the fields $A$ and $c$, along with their corresponding antifields $A^+$ and $c^+$ are covariant. As such, they transform tensorially under the change of coordinates on $M$. Let $M^i_j(x)$ be the Jacobian  matrix of such transformation, and $M^a_b(x)$ the induced Jacobian matrix on $E$. Then the components of $A$ transform into:
\begin{equation}
\widetilde{A}^a=M^a_b A^b\,,
\end{equation}
and the same way for the other fields.

Knowing the transformation of these fields, along with the transformation of $X$, and taking into account that the symplectic form \eqref{sym_form} has to be invariant under changes of coordinates, one can find the transformation of $X^+$:
\begin{equation}
\widetilde{X}^+_i=(M^{-1})^j_i X^+_j-(M^{-1})^c_b\, \partial_\mu M^a_c (-A^+_a\wedge A^b+c^+_a c^b)\,.
\end{equation}
With this transformation in hand, it is easy to check the covariance of the BV action. However, in order to make this covariance manifest, one needs to covariantize $X^+$ first:
\begin{equation}
{X}^{+\nabla}_i=X^+_i-{\omega}^{a}_{bi}\left(-A^+_a\wedge A^b+c^+_a c^b\right)\,.
\end{equation}
which now transforms as a tensor. In terms of this new field, the BV action takes the form:
\begin{eqnarray}
\mathcal{S}_{BV} &=& -\int_{\S}\left(||F||^{2}+\bigg\langle\theta(A),\dd X+\frac 12 \rho(A)\bigg\rangle\right)-\int_{\widehat{\S}}X^{\ast}H\nn\\[4pt]
&& +\int_\Sigma \left(\left\langle {X}^{+\nabla},\rho(c)\right\rangle-\frac{1}{4}(T^+ +T^-) (c^{+},c,c)+\langle\phi,\ast F\rangle(A^{+},c)\right)\nn\\[4pt]
&& -\frac{1}{2}\int_\Sigma \left(\left((D^+ +D^-)c\right)(A^+)-(T^{+}+T^{-}) (A^{+},A,c)\right)\nn\\[4pt]
&& -\frac{1}{8}\int_\Sigma \left(\left\langle S^+(A^+,c,c),\mathcal{G}^{-1}_-(A)\right\rangle+\left\langle S^-(A^+,c,c),\mathcal{G}^{-1}_+(A)\right\rangle\right)\nn\\[4pt]
&& -\frac{1}{8}\int_\Sigma\left(\left\langle S^+(A^+,c,c),\mathcal{G}^{-1}_-(\ast A)\right\rangle-\left\langle S^-(A^+,c,c),\mathcal{G}^{-1}_+(\ast A)\right\rangle\right)\,,
\end{eqnarray}
where pullbacks by $X$ are assumed.

\section{Conclusion}

Even though it has been known for a while how Dirac sigma models arise from gauging of the nonlinear sigma models \cite{Chatzistavrakidis:2016jci}, we have shown that this gauging can be taken to be quite more general. By including the possibility of nonminimal coupling in the metric sector, it turns out that in many cases this turns out to be just a simple redefinition of minimal coupling. Despite that, it would be interesting to understand whether the remaining cases can be turned to minimal coupling as well or they give rise to a different geometrical structure.

Finally, we have constructed the classical BV action for the Dirac sigma models. Since the target space for Dirac sigma models is not a QP manifold, it is not possible to rely on AKSZ construction for this. Instead, this has been achieved by expanding the BV action in terms of the antifield number and then solving the classical master equation directly.

\paragraph{Acknowledgements.} We would like to thank A. Chatzistavrakidis, L. Jonke, Z. Kökényesi and T. Strobl for useful discussions and contributions. This work was supported by the Croatian Science Foundation Project ``New Geometries for Gravity and Spacetime'' (IP-2018-01-7615).

\end{document}